# Developing atomistic glass models using potential-free Monte Carlo method: From simple to complex structures

Shakti Singh and Sharat Chandra

Materials Science Group, Indira Gandhi Centre for Atomic Research, HBNI, Kalpakkam, TN, 603102 India

E-mail: S.S: shaktisinghstephen@gmail.com, S.C: sharat.c@gmail.com

**Abstract**

*We propose here a method to generate random networked amorphous structure using only readily available short-range properties like bond lengths, bond angles and connectivity of the constituents. This method is a variant of Monte-Carlo (MC) method wherein the basic constituents of an amorphous network i.e. rigid polyhedral units are connected randomly obeying certain steric constraints. The algorithm is designed to reproduce the medium-range order universally observed in glasses. The method somewhat resembles the reverse MC (RMC) method where a random move of an atom inside a box is accepted or rejected depending upon whether it decreases or increases the deviation from the experimentally observed features. However unlike RMC, this method does not demand large experimental sets of scattering data which in most cases is **a priori** not available for glasses. It rather relies on the stochasticity of MC method to produce glassy structures. The algorithm is first validated against SiO$_2$ glass structure by comparing with the available structures from other methods and experimental data. The method is then extended for developing more complex Iron Phosphate Glass (IPG) structures and a comparison with existing models of IPG developed using quench-from-melt scheme implemented in classical Molecular Dynamics (MD) reveals that the method is extensible to complex glasses also. This study addresses the often-neglected issue of non-availability of correct starting structures in simulating glasses using MD or Density Functional Theory (DFT).*

## 1. Introduction

The endeavour to model atomistic structures of glasses has been around since the time Zachariasen[1] gave insight into the structure of glasses. Atomistic structure of glass holds the key to explanation of microscopic origins of a wide range of phenomena like their structural stability, superior glass forming ability (GFA) of certain compositions of elements over the other, superior chemical durability of certain glass compositions over others etc. Moreover atomistic structures are useful in radiation cascade studies for suitability in vitrification of waste generated from reprocessing of spent nuclear fuel and development of interatomic potentials for MD simulation of glasses. Yet a complete knowledge of atomic coordinates inside a glass has remained elusive because most experimental probes give reliable information only upto short-range since the intermediate-range and long-range information is averaged out due to randomness in structure at these length scales. This shortcoming of experiments to fully predict the atomistic structure of glass is a well-known fact and warrants for the need of computational models of glasses to study them[2]. At present, structural models of glasses that can reproduce all experimentally known properties are still lacking[3].

Initial attempts at modelling glassy structures were made using handmade models[4] which gave way to computer generated models with the advent of computers. Now with fast cluster computing facilities and parallel codes of ab-initio calculations, it is possible to develop atomistic models of any system by simulating the experimental melt-quench technique albeit at very fast time scale (picoseconds in DFT to nanoseconds in MD). The last two decades have witnessed unprecedented rise in efforts to develop glassy or in general amorphous structures based on computational methods such as MC methods [5–7], MD simulation [8], DFT based method [9] also called ab-initio MD (AIMD) or first principle MD (FPMD) and hybrid methods[10,11]. The authors still think for a need to develop the presented algorithm due to the following reasons:

(i) **Absence of correct starting structures in the current methods**: The most popular method for reaching a large size glassy structure is MD based melt-quench method. Using the present computational resources, one can simulate the dynamics of this process but at very fast time scale (picoseconds in DFT to nanoseconds in classical MD). The shortcoming of such simulation is that the cooling rates (CRs) employed are extremely fast (~$10^{13}$ K/s) as compared to experimental CRs (typically 10 K/s for air quenching). Due to this the final glassy structure inherits structural similarities of the starting structure i.e. the pre-melt-quench structure. Most studies start with a crystal structure with a similar composition and connections of constituent units as found in glass, leading to melt-quench structures which inherit crystalline properties of the starting structure. If starting structures are fairly glassy (in the sense that they obey most of the experimentally known short-range and medium-range properties of that glass and at the same time lack long-range order) then the final structures could be more realistic. Such starting structures can also expedite the geometry relaxation jobs in AIMD based development of glasses or in RMC methods. The fundamental reason why this should work can be seen from the potential energy landscape (PEL). By prebuilding the features known experimentally into the developed models we are ensuring that the search algorithm stays near the metastable glass valley in the PEL and hence quicker convergence can be guaranteed.

(ii) **The computational time taken by the current methods**: Classical MD can generate very large structure (upto millions of atoms) within reasonable time of few days through the melt-quench method, owing to solvation of less demanding Newton's equation of motion. However the structures are as good as the interatomic potentials used and, in most cases, good interatomic potentials are *a priori* not available. The AIMD methods takes huge time as it solves more complex Schrodinger equations of motion for electronic minimisation and Newton's equation for ionic minimisation. This limits the structures to only hundreds of atoms which leads to incorrect properties due to absence of long-range disorder. Having fairly glassy structures as starting structures can reduce the search time for optimized glassy structure in both classical MD as well as AIMD methods since we already started in the vicinity of the metastable valley in PEL.

The remedy proposed here is as follows. Since it is computationally intensive to go ab-initio in glass model development, so one can initially model any glass structure using information which has been already well established from experiments i.e. short-range properties like nearest neighbour bond length, bond angle distribution and coordination number. Moreover, Raman scattering based experiments have been able to tell us about the defects present in realistic glass structures like the percentages of under-coordinated and over-coordinated atoms on account of edge-sharing and face-sharing among polyhedral units. So, these defects can also be introduced randomly during the MC modelling to produce realistic glass structures. For correctly reproducing most of the short, medium, long-range properties the presented method relies on the stochasticity of the MC method in mimicking the randomness in glass networks. Short range order is fed into our algorithm and intermediate range is designed by our algorithm. The structures use periodic boundary conditions (PBC) and hence they can further be used in DFT/MD simulations to determine their electronic/mechanical properties. The models thus produced are not force-relaxed due to the potential-free nature of our MC method. Hence, they have to be relaxed to the minimum energy configurations using the DFT/MD methods or they can be directly used as starting structures in MD melt-quench method to produce more realistic glass models. The initial modelling from our code helps to bring the system closer to the actual metastable valley in the PEL and subsequent force relaxation can take it to the valley itself.

The rest of the paper is organized as follows: Section 2 discusses the computational method designed in this paper to model glassy networks with subsections discussing the method overview, implementation details and the algorithm respectively. Section 3 discusses the results of implementing this algorithm for $SiO_2$ glass (bench-marking) and subsequent extension for complex system of IPG.

## 2. Computational Method

### 2.1 Method overview

The method described here is broadly based on an old technique of amorphous structure modelling in which starting from a seed, a dense random network is grown by adding a basic repetitive unit according to some aggregation criteria suitable for the specific system[12]. For the sake of classification, this method can be included in the "decorate and relax" scheme in the information paradigm of developing glassy structures as described in the review by D. A. Drabold[3]. Force relaxation of the final structures becomes a necessary step in such methods due to their potential free nature.

The structure of glass is composed of the *short-range structure* decided predominantly by the chemical bonding involved, which dictates the basic geometry and connectivity in glasses which is repeated throughout in the 3-dimensional structure. This information is readily available usually from neutron scattering experiments and is used to define the basic repetitive unit in our method.

At a longer length scale is the complicated *intermediate-range structure* decided mainly by the kinetics of glass formation. The difficulty of most experimental methods to quantify this IRO in glasses is because of the randomness involved in the way the basic constituent units defined in SR are connected with each other. However there is a growing consensus that the first sharp diffraction peak (FSDP) observed universally in the diffraction pattern of glasses is due to the characteristic intermediate-range clusters found in all glasses[13]. For reproducing this in our models, a special algorithm is designed which we called search-join-place algorithm discussed in the implementation details.

At still longer length scale is the *long-range structure* which has no well-defined order and defines the macroscopic properties of the glass such as density, void fraction and specific heat. Most important long-range structure dependent property from modelling point of view is density. It is available from experiments and here it is used to tell the program about the number of atoms and size of simulation box. More detailed discussion on structure of glasses are available in the literature.[14–18]

To start with, we have modelled the simplest among glass structures i.e. silica glass structure. The short-range structure of $SiO_2$ glass is known from XRD or Raman scattering experiments. It is known to be composed of well-defined tetrahedral units of $[SiO_4]^{4-}$ which has very less deviation in the Si-O bond-length of 1.59 Å as well as in O-Si-O bond angle of 109.5°[19]. So, this aspect of glass is given as an input to our algorithm. Next, these polyhedral units or *motifs* are connected with each other at random values of the bridging bond angle i.e. angle Si-O-Si (see Figure 4(b)). It is this orientational degree of freedom (dof) of the units which is randomly sampled adhering to steric constraints to obtain the long-range disorder found in glassy structures. Apart from this we have also designed an algorithm to reproduce the intermediate-range structure of glasses.

## 2.2 Implementation details

**Constraints**: The motifs placed inside the simulation box must always adhere to the steric constraints between constituent atoms. For this purpose, successive motifs are placed so that all atoms have these constraints satisfied. The minimum cut-offs for oxygen and silicon atoms belonging to different motifs are $O_i$-$O_j$ = 2.3 Å, $Si_i$-$Si_j$ = 2.7 Å where i and j subscripts belong to two different polyhedral motifs. Similarly, the bond-angle Si-O-Si is allowed to vary between 120° and 180° and rest of the values are prohibited. These values are consistent with experimental results [19,20].

**Rodrigues rotation Formula**: For 3D rotation of the motifs inside the simulation box we have used the Rodrigues rotation formula [21] to rotate a vector in space given an axis (**K**≡($n_1$,$n_2$,$n_3$)) and an angle of rotation (θ). Axis for this purpose is taken to be in Si-O direction where O is the oxygen on which the next motif is planned and θ is the bridging bond angle Si-O-Si. Basic steps involved in the rotation of a motif that is going to be placed at oxygen(**r**) are explained in Figure 1. In matrix notation the formula is given below[22]:

$$R = \begin{bmatrix} \cos\theta + (1-\cos\theta)n_1^2 & -n_3\sin\theta + (1-\cos\theta)n_1n_2 & n_2\sin\theta + (1-\cos\theta)n_1n_3 \\ n_3\sin\theta + (1-\cos\theta)n_1n_2 & \cos\theta + (1-\cos\theta)n_2^2 & -n_1\sin\theta + (1-\cos\theta)n_2n_3 \\ -n_2\sin\theta + (1-\cos\theta)n_1n_3 & n_1\sin\theta + (1-\cos\theta)n_2n_3 & \cos\theta + (1-\cos\theta)n_3^2 \end{bmatrix}$$

The matrix operated on a column vector produces the coordinates of rotated vector.

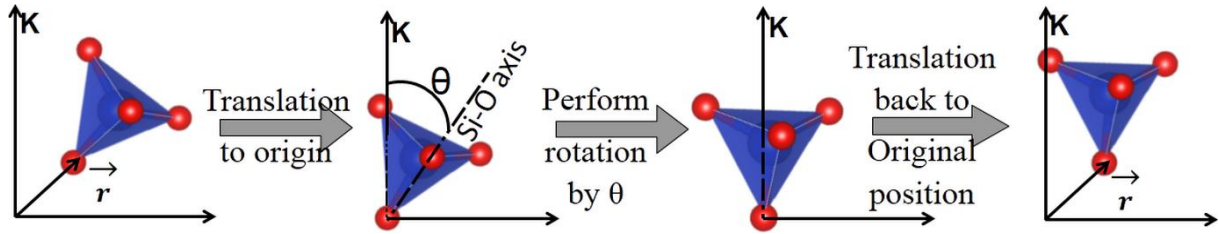

*Figure 1: Scheme of performing rotation at arbitrary point r*

**PBC in structures:** This aspect of the atomistic structures is very important if they are to be used further in atomistic level codes that use methods like DFT or MD. PBC in simulation cell allow these codes to predict the bulk properties of materials. Although the use of PBC in predicting properties of amorphous systems may not be desirable since they lack long-range order but any such shortcoming can be overcome to some extent by using a larger simulation cell. Also note that the PBC as implemented in our models work only when all the three axial angles are 90° i.e for orthorhombic systems.

Periodic boundary conditions should be kept in consideration in any atomistic modelling at two levels. Firstly while calculating distances between atoms. All the distances in the program are calculated respecting PBC i.e. if the difference between the x, y, z coordinates of atom A and atom B is less than or equal to half

### *PBC Wrapping equations*

$$X_f = X_i - X_{max} * int\left(\frac{X_i}{X_{max}}\right), \qquad for\ X_i \geq 0$$

$$= X_{max} + X_i - X_{max} * int\left(\frac{X_i}{X_{max}}\right), \qquad for\ X_i < 0$$

of the respective dimension of the cube along that axis, then the original distance AB is minimum distance between those points. But for the other case, where the difference exceeds half of cell dimension along that axis, atom A is translated by the cell dimension along that axis and distance A′B between image of atom A i.e. A′ and atom B is considered as the shortest distance. A pictorial aid for understanding this is given in Figure 2(a). Distance calculation is a central part of the program since all the atoms should follow the constraint cut-offs placed for ensuring that no two atoms come closer than a cut-off distance.

Secondly while generating the model, any atom moving out from any boundary wall is wrapped inside from the opposite wall using the PBC wrapping equations. During such wrapping, the PBC image may end up getting too close to oxygen of tetrahedral unit already placed. This warrants the constraint cut-offs to be adhered in consistency with PBC. This is shown in Figure 2(b). Although periodicity in amorphous structures is not present, still for the purpose of modelling it becomes a necessary evil. The effect of periodicity on the predicted properties is also an area of study in amorphous modelling.

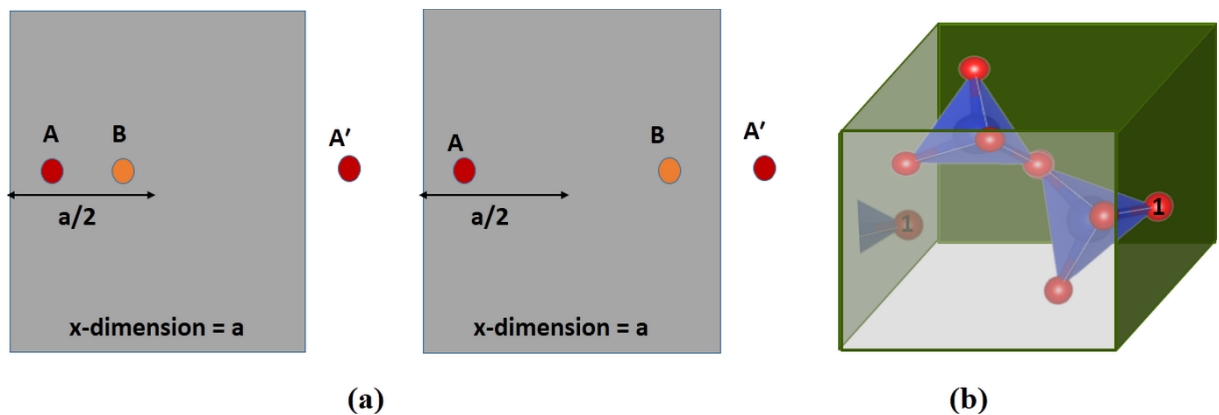

*Figure 2: PBC (a) **Distance Calculation**: left shows the case where distance AB < a/2, so AB is the shortest distance between A and B. Right shows case where AB > a/2, so A'B is the shortest distance rather than AB. Has to be done for each of the x, y and z coordinates of the distance vector and appropriately the distance vector has to be modified. (b) Tetrahedral unit placed near the right boundary has oxygen1 (in red) outside the simulation box, hence it enters inside the box from left face according to PBC wrapping equation, but the PBC image of oxygen1 may end up getting too close to oxygen of tetrahedral unit already placed. This warrants the constraint cut-offs to be adhered in consistency with PBC*

Described next is an important part of the algorithm without which the experimental density and stoichiometry can never be obtained in an atomistic model.

**Search-join-place algorithm**: The motifs placed successively should form closed ring like structure instead of forming a dendrite like structure. Such ring like structures are characteristic of the intermediate range order found in most glasses[23] and also hold the key for FSDP seen universally in structure factor of glasses. Therefore, while generating the model from initial seed, the algorithm continuously searches for unconnected (bonded to only one Si atom) nearby oxygen atoms of already placed motifs and utilises them as either edge or face of forthcoming motif. This ensures appropriate connectivity in model. Upon using the nearby oxygen as edge or face of newly placed tetrahedral motif, the program tries various possibilities to put the complete motif. But if the steric constraints do not allow the placement of new tetrahedral unit, then the new tetrahedral unit cannot be placed. At such places Si is put without fulfilling its 4 valency. Such under-coordinated Si atoms having a dangling bond are found even in realistic glasses in varied concentration depending upon glass preparation conditions. The consistency of this algorithm with PBC is also ensured. We named it as the search-join-place algorithm after the steps that are involved in it.

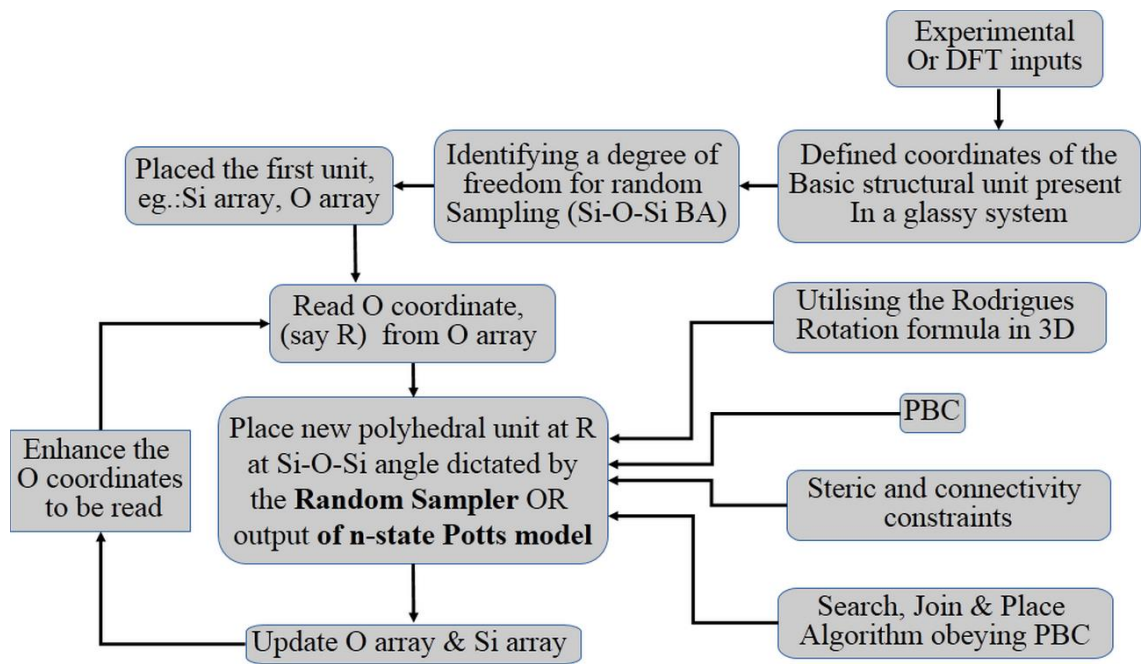

*Figure 3: Flow-Chart of the algorithm used to develop glass structure*

### 2.3  Algorithm

Figure 3 gives the flowchart of the algorithm used to develop $SiO_2$ glass structure. Starting with the short-range information available from experiments or ab-initio studies like DFT, one defines the coordinates of the smallest repeatable motif which for the case of $SiO_2$ is the $[SiO_4]^{4-}$ tetrahedral unit. Then we have the well-established observation that disorder in glass structure can be attributed to the randomness involved in the way these motifs connect with each other[14]. Therefore, we identify an orientational dof of the structure that can be randomised to obtain a random-network. For the case of $SiO_2$, it is the Si-O-Si angle also called the bridging bond-angle or the inter-tetrahedral angle. Thereafter, an orthogonal simulation box is defined and the first motif is placed at a random position inside it. Computationally, this amounts to assigning coordinate positions of oxygen and silicon atoms to an array. Now the oxygen coordinates are successively read from this array and new motifs are placed at each vertex oxygen at Si-O-Si bond angle dictated by a 'random sampler'. The array is updated each time a new unit is placed. We have initially tried to use the output of n-state Potts model[24] to randomly sample the orientational dof in glass due to the success of this model in reproducing dynamic behaviour of glass melts[25]. For this purpose, the spin orientations in Potts model can be correlated to the orientation of polyhedral units in glasses. However, because of the discreteness of spin orientations in Potts model, we concluded that for modelling glasses it is more appropriate to use a more continuous random sampler available in most programming languages. We have used Python for programming the code, and hence its random sampler is used which is based on Mersenne-twister algorithm[26]. The random rotation of tetrahedral unit about the Si-O axis is performed using the Rodrigues rotation formula[21] already mentioned. Only those units are placed which are permitted by the constraints. The search, join and place algorithm is necessary to reproduce the ring topology of the glassy structure. Lastly, the structures are generated conforming to PBC which is very important to correctly reproduce the experimental density, stoichiometry and void distribution.

Figure 4 shows the schematic of developing glassy structures of $SiO_2$ by our method within a cubic simulation box of edge-length 36 Å as an example. The final structure developed is also shown in Figure 4*(d)*. For visualisation, VESTA[27] software is used.

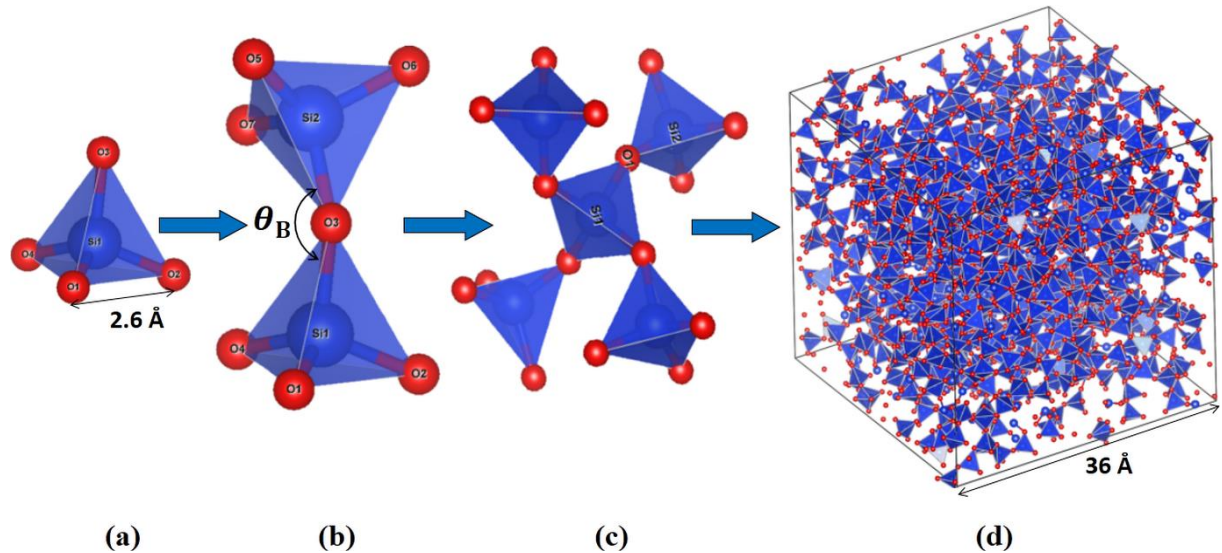

(a)    (b)    (c)    (d)

*Figure 4: Successive stages involved in developing silica glass structure (a) the input tetrahedral unit (b) the dimer unit obtained by placing another tetrahedral unit at one of the corner oxygen atoms such that the bridging bond angle Si-O-Si (labelled $\theta_B$) is randomly sampled (c) all four oxygen atoms of central motif bridged to obtain a connectivity of 2 for the oxygen atoms (d) Developed structure of silica glass consistent with PBC, having density of 2.26 g/cc (experimental density of $SiO_2$ glass is 2.20 g/cc). For visualisation VESTA software[27] is used.*

## 3. <u>Results and Discussion</u>

## 3.1 Algorithm Validation: Case study on $SiO_2$ glass

The structure developed for silica glass is then benchmarked against the rich literature available for silica glass [28–31]. We discuss here the properties of one of the developed structures having 2952 atoms in cubic box of edge length 36 Å. Existing atomistic models used for comparing the structural properties were developed using DFT (162 atoms in triclinic cell of dimension 13.72, 12.69, 14.16 and angles 91°, 93.6°, 90°) and RMC method (3000 atoms in cubic box of 36 Å) which we will call the DFT model[31] and the RMC model[32] respectively. Apart from these theoretical models, comparison is also done with available experimental data. The analysis is done in a systematic manner starting from the short-range properties to intermediate-range and finally the long-range properties.

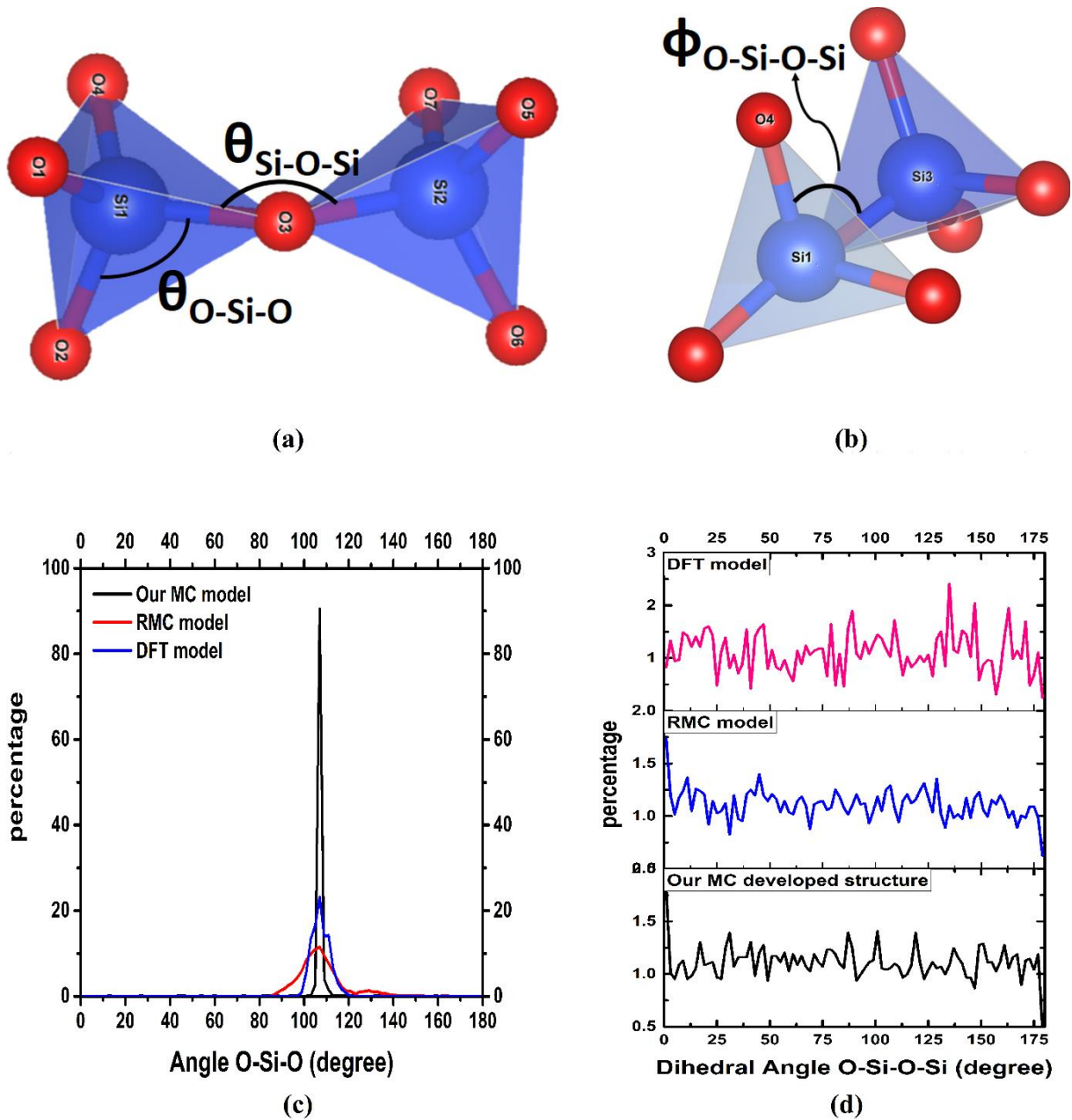

*Figure 5: Showing the connections among basic building blocks of $SiO_2$ glass to define (a) the intra and inter-tetrahedral bond-angle (O-Si-O and Si-O-Si respectively) (b) the dihedral angle $O_4$-$Si_1$-$O_2$-$Si_2$ when seen along $Si_1$-$O_2$ axis (hence the bridging oxygen $O_2$ is not visible) (c) graph showing the distribution of intra-tetrahedral bond angle and (d) distribution of dihedral bond angle O-Si-O-Si in our model and other theoretical models of $SiO_2$*

**Short-range properties:**

The structural properties of generated model at short length scales used to characterize the short-range structure are: radial distribution function (RDF) for nearest neighbour and next nearest neighbour bond distances, partial-pair distribution functions (PDFs), neutron structure factor S(Q), inter-tetrahedral (Si-

O-Si), intra-tetrahedral (O-Si-O) and dihedral (O-Si-O-Si) bond angles distributions and coordination number analysis.

In Figure 5, we have shown in (a) the inter-tetrahedral and intra-tetrahedral bond angle and in (b) the dihedral angle in silica glass structure. In (c) the intra-tetrahedral bond angle distribution and in (d) the dihedral bond angle distribution in the DFT, RMC and MC based models is shown. The intra-tetrahedral bond angle distribution O-Si-O is found to peak at ~109° in the three models discussed here. Since the MC model is not force relaxed, the peak in its O-Si-O angle distribution is sharper than other two models. This was expected since we used rigid tetrahedral units as the basic building block. This peak is expected to get slightly broader after a force-relaxation. The dihedral angle in all the models is distributed throughout the range 0° to 180° which should be the case in a random network structure.

The RDF comparison in Figure 6(a) clearly show a good agreement between our model and existing theoretical models as well as the experimental RDF[33]. The first peak at 1.59 Å is due to the Si-O bond length followed by second peak at 2.6 Å due to the O-O distance in tetrahedral motif. The third peak (not clearly visible due to scale) at about 3.05 Å is due to the Si-Si distance between neighbouring tetrahedral motifs. This peak was easily identified in partial PDF graphs shown in Figure 6(b). The comparison of partial PDF shows more clearly the presence of the first three peaks in total RDF. Subsequent peaks in RDF of any glass are suppressed due random structure present in them beyond this range and hence cannot be ascertained. However, a glassy structure warrants that RDF should approach 1 sufficiently faster with radial distance and this happens at ~6 Å onwards in our structure.

In Figure 6(c), neutron structure factor $S(Q)$ is compared. The experimental data is also presented for comparison[34]. The presence of FSDP at small wave vector Q (shown by dotted line at ~1.5 Å$^{-1}$) in all the curves is crucial confirmation of intermediate range order in a glassy structure which will be discussed later. In addition the DFT model has peaks even before the FSDP, which is due to the periodicity of the structures produced via DFT and is undesirable. Apart from this, all the $S(Q)$ curves show one to one correspondence among all the peaks, validating the structures produced by our code.

Another important measure to quantify order among disorder structures is the intertetrahedral bond angle Si-O-Si distribution shown in Figure 6(d). This angle is experimentally found to vary between 120° to 180°, such that the distribution has peak around 144°-147° and is asymmetric with lower angles favoured more[20,35–37]. Our MC developed model and RMC model show this peak at ~140° which is slightly on the lower side as compared to the experimental value. The DFT model correctly reproduces this peak.

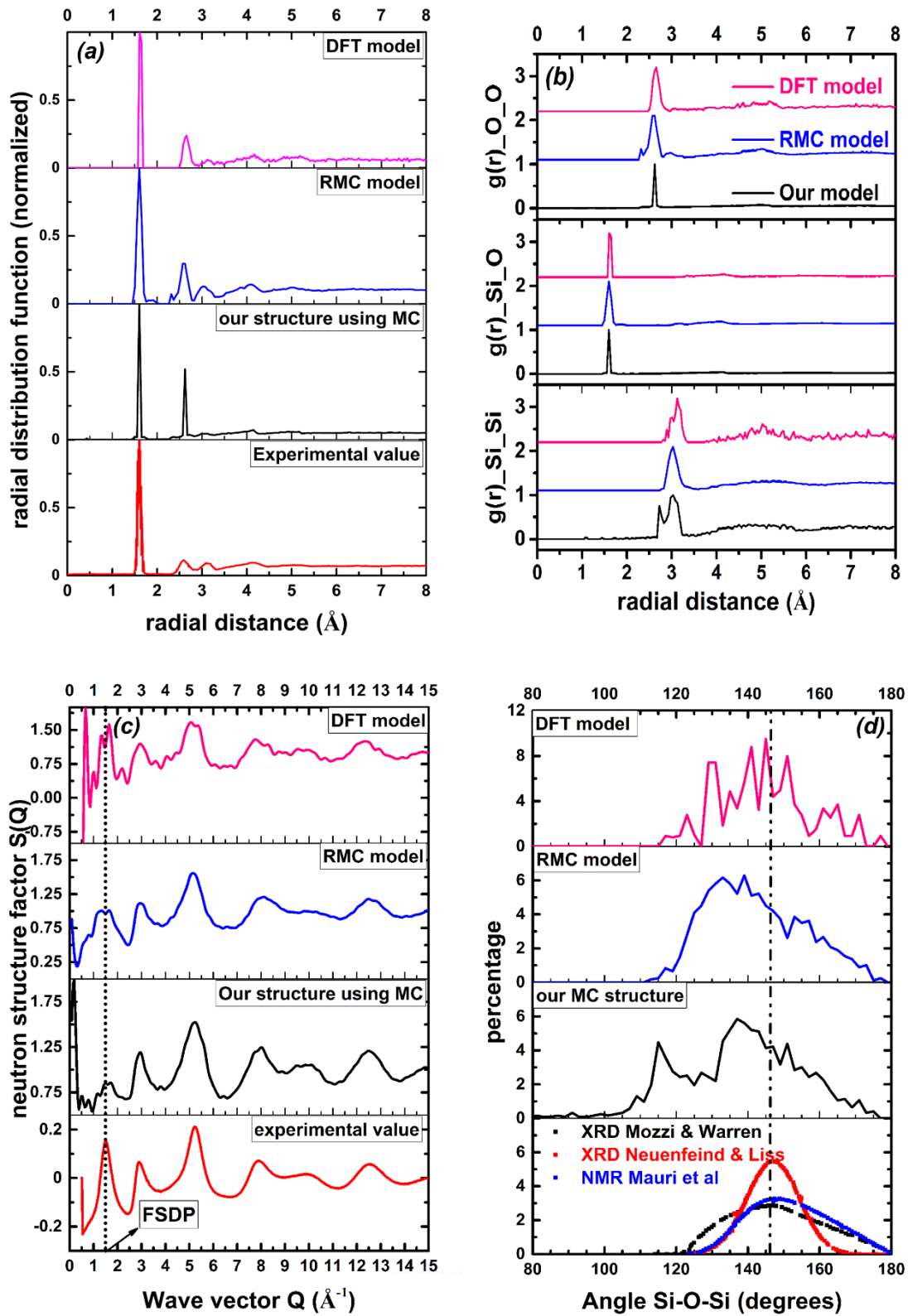

Figure 6: Graphs for analysing the Short-range structure of $SiO_2$ glass showing (a) radial distribution function (b) partial pair distribution function (c) neutron structure factor S(Q) (d)

*inter-tetrahedral bond angle Si-O-Si distribution. The structural properties of our MC-based model (black) are compared with that of existing theoretical models (DFT model (pink) and RMC model (blue)) and experimental values (red) wherever available*

We also analysed coordination numbers for our structure and compared them with the RMC and DFT model. These values are obtained using R.I.N.G.S. package [38] and are tabulated in Table 1 and Table 2. The DFT model is a perfectly coordinated structure, whereas the RMC model does exhibit coordination defects which are present in realistic glass as well and the concentration of which depends on glass preparation history. Our MC model has more concentration of these defects which is also observed in the density value being less (2.07 g/cc) than experimental value (2.2 g/cc). We are extending the code to incorporate user-defined concentration of defects in the modelled structure.

*Table 1: Showing nearest-neighbour coordination number (CN) analysis for the configuration around Si atoms (perfect value of CN=4)*

| Si coordination | % distribution in DFT model | % distribution in RMC model | % distribution in our model |
|---|---|---|---|
| 1 | 0 | 0.2 | 1.01 |
| 2 | 0 | 4.6 | 13.6 |
| 3 | 0 | 16.5 | 2.8 |
| 4 | 100 | 78.7 | 82.4 |
| **Average CN** | **4** | **3.74** | **3.66** |

*Table 2: Showing nearest-neighbour coordination number (CN) analysis for the configuration around O atoms (perfect value of CN=2)*

| O Coordination | % distribution in DFT model | % distribution in RMC model | % distribution in our model |
|---|---|---|---|
| 0 | 0 | 0.35 | 0 |
| 1 | 0 | 12.5 | 40.8 |
| 2 | 100 | 87.1 | 58.7 |
| 3 | 0 | 0.05 | 0.3 |
| **Average CN** | **2** | **1.87** | **1.59** |

**Intermediate-range properties:**

**FSDP**: After characterising the short range structure, next step is to analyse intermediate range structure. Figure 6(b) shows the FSDP in neutron structure factor shown by dotted line at Q value of ~1.5 Å$^{-1}$. The presence of this peak in all the models studied in this paper confirms the presence of medium range order in all the structure. This peak has been suggested to correspond to largest repeatable unit in the network.[39] We did a quantitative analysis of the FWHM of this peak by fitting it with Lorentzian function, for different models to estimate the coherence length corresponding to this peak (coherence length L=7.7/FWHM [40,41]). In SiO$_2$ glass we have rings of type -Si-O-Si-O- and unlike Beta-crystobalite, we have a distribution of ring sizes around a peak value which occurs at 12-node ring size. The peak in FSDP may be related to the intermediate range order that these rings impart to glass as observed in other studies as well[42]. The FSDP Q value, FWHM and the corresponding coherence length is reported for the three models discussed in this paper together with experimentally observed values in Table 3. The FSDP Q value, its FWHM and coherence length of our MC model are in good agreement with the values obtained experimentally and from DFT model. The RMC model has FWHM of ~1.48Å$^{-1}$ which translates to coherence length of 5.20Å which implies more disorder in the medium range structure of this model than experimentally observed. For comparison the coherence length of crystalline form of SiO$_2$ having similar density (2.2 g/cc) as silica glass, i.e. β-crystobalite form is also reported. The coherence length for glassy models should be less than that for crystalline form due to disorder present in the glass, and this is observed for the models discussed in this study, albeit the RMC model has far too less value, which is undesirable for a good structure.

*Table 3: FSDP in neutron structure factor, its FWHM and coherence length for various models of a-SiO$_2$ discussed in this study, experimental values and for crystalline SiO$_2$ (β-Crystobalite)*

| Structural model | FSDP Peak location Q (in Å$^{-1}$) | FWHM (in Å$^{-1}$) | Coherence length (in Å) |
|---|---|---|---|
| DFT model | 1.53 | 0.55 | 14.0 |
| RMC model | 1.47 | 1.48 | 5.20 |
| Our MC model | 1.62 | 0.64 | 12.03 |
| Experimental | 1.53 | 0.53 | 14.53 |
| β-Crystobalite | 1.48 | 0.48 | 16.04 |

**Rings**: The structure of glass at intermediate length scale is full of facts unexplored experimentally since none of the probe available at present is able to measure a property which can be correlated to it. Nevertheless, there is increasing consensus among modelling experts that Topological disorder of the glass network can be quantified in terms of the rings size distribution. So, there are lots of studies on rings statistics of structural models to characterise the IRO structure.[23,43,44,42] An irreducible ring in the context of glasses is defined as smallest closed chains/paths formed by the bridging oxygen atoms and network formers, starting and ending at the same atomic site in the glassy network. In counting such rings, care should be taken to omit those rings that can be decomposed into two or more smaller rings. There are two ways of specifying a ring size. We can either count all the atoms in the path of a ring or only the network-formers (eg. Si for $SiO_2$) can be counted. We will stick to the former convention. In this paper, the IRO is studied using the R.I.N.G.S. package[38]. Out of the many possible different definitions of a ring inside a structure, we will stick to Guttman's shortest path criterion[45] to define a ring. According to it, starting from a node, a ring corresponds to the shortest path which comes back to the given node from one of its nearest neighbours. Figure 7(a) shows a 12-atom ring present in our structure and in (b) a snapshot of various rings present in our structure. In crystalline forms of $SiO_2$, one finds the presence of only single sized rings. For example in coesite form of silica only 8-node rings are found, keatite has 10-node, rest of the forms like quartz, cristobalite and tridymite have 12-node rings only. In contrast, silica glass structure possess a distribution of ring sizes from 6-node to 16-node. Shown in Figure 7(c) is the distribution of rings in our model as compared with the RMC model and the DFT model. As we can see the structures have peak near 10-12 atom ring. The graph dips on either side of this ring size. This topological behaviour is confirmed by independent studies on ring statistics in $SiO_2$ system [46] [42]. We observe here that the RMC model is best in terms of reproducing the ring statistics in silica glass. The DFT model shows mainly 10 and 12 membered rings but usually we have fair amount of 6, 8, 14 and 16 node rings as well. Our MC model has fairly high percentage of larger rings (16, 18, and 20) which is inconsistent with usual observation.

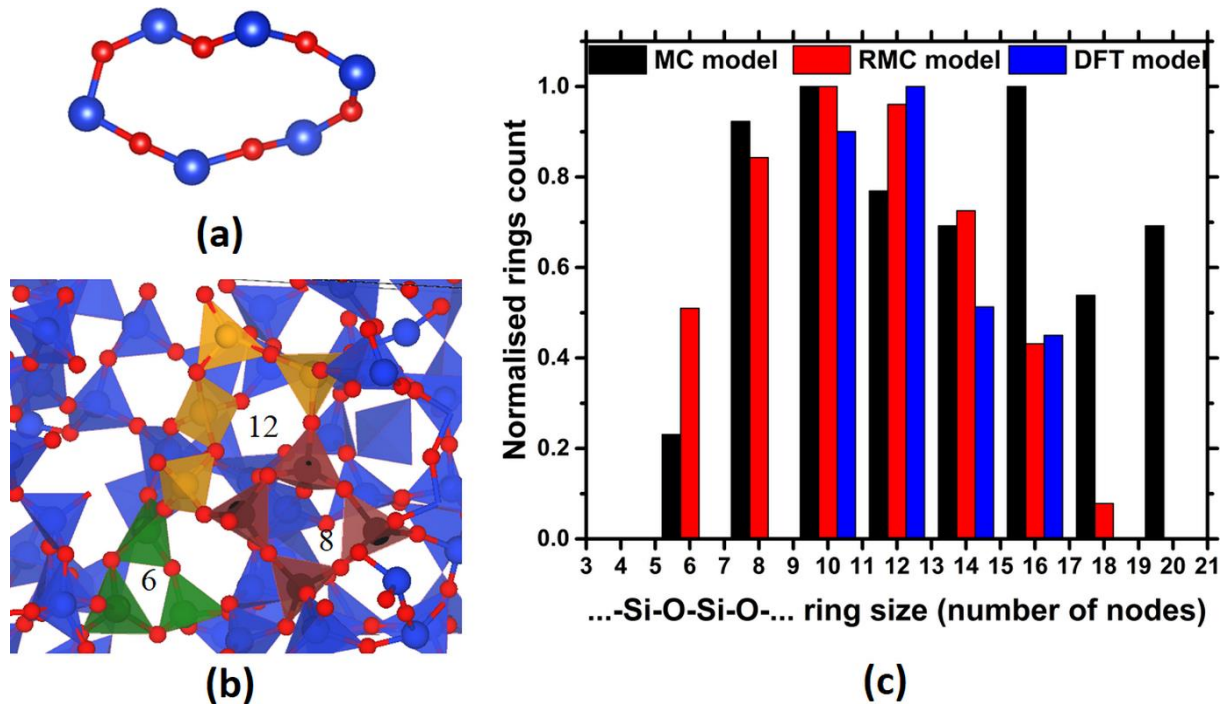

Figure 7: Ring analysis (a) shows the image of a 12-atom ring present in our structure (b) shows the presence of various ring size in our system. Images are taken using VESTA software[27] (c)

*shows the distribution of rings with respect to their size in our structure as compared to the RMC and DFT model*

**Long-range properties:**

For validating the long-range structure of our model, we compared the density and void fraction of our model with other theoretical models and experimental values, reported in Table 4. Although the density values are quite close to the experimental value, the void fraction of all the models is found to be consistently different from void fraction obtained using positron annihilation lifetime spectroscopy[47]. The reason for this discrepancy lies in the difference in the void space that positrons probe versus the void space calculated using theoretical methods. In void calculation of theoretical models, we have taken the Van der Waal's radius for atoms, which may not exactly mimic the realistic case.

*Table 4: Comparison of density and void fraction of our model with existing theoretical models (DFT and RMC) and experimental values*

| Structural Model | Density (g/cc) | Void Fraction |
|---|---|---|
| **DFT model** [31] | 2.19 | 0.31 |
| **RMC model** [32] | 2.14 | 0.33 |
| **Our model** | 2.07 | 0.30 |
| **Experimental** [47] | 2.20 | 0.18 |

The agreement between structural properties from our developed structure with that from RMC model, DFT model and experiments shows that the presented method can indeed produce good structural models using minimum information about the system. Our code for $SiO_2$ produces glass structures with coordination defects which are found to be present in realistic glass as well. Moreover our code can also be used to produce structures satisfying different orientational constraints for the constituent motifs. Effect of such strains on glass properties has been an important research topic[30]. However the real success of this method lies in its extensibility to more complex glass structure. So the code is redesigned to develop Iron Phosphate glass (IPG) models.

## 3.2 Extension of the algorithm for complex glasses: Modelling the Iron Phosphate Glass (IPG)

The algorithm is extended for $40Fe_2O_3:60P_2O_5$ (by weight) composition of Iron Phosphate glass (henceforth this composition is referred as IPG). This glass is currently being studied extensively for vitrification of waste generated during reprocessing of fast reactor spent fuel. Its suitability for this purpose is argued nicely in references [48][49]. There were attempts to provide atomistic models of IPG starting from DFT studies[50,51] to some recent MD based studies[52–54]. Yet due to the deficiencies associated with these methods (as discussed earlier), we adopted this study to provide good atomistic models for simulation studies of these glasses which are presently lacking.

IPG glass modelling poses several challenges.

(1) The motifs involved now are phosphate tetrahedral units $(PO_4)^{3-}$, $(FeO_6)^{9-}$ octahedral and $(FeO_4)^{5-}$ tetrahedral unit (having Fe in 3+ oxidation state), $(FeO_4)^{6-}$ tetrahedral units or trigonal prism units (having Fe in 2+ oxidation state)[55]. The presence of iron in 2+ oxidation state depends on the experimental conditions of manufacturing the iron phosphate glass like melting temperature, atmosphere, time and quenching technique. In the literature, compositions of this glass having $Fe^{2+}$/Fe in range 0% to 20% are studied. In the program we have developed, concentration of various species can be specified in the beginning. The presence of these motifs in this composition has been studied experimentally through spectroscopic techniques as well as computational techniques like DFT and MD in references [51][50][54][53][52]. Figure 8 shows two crystal structures of iron-phosphate showing dominant coordination geometries and connection types of polyhedra as observed in IPG.

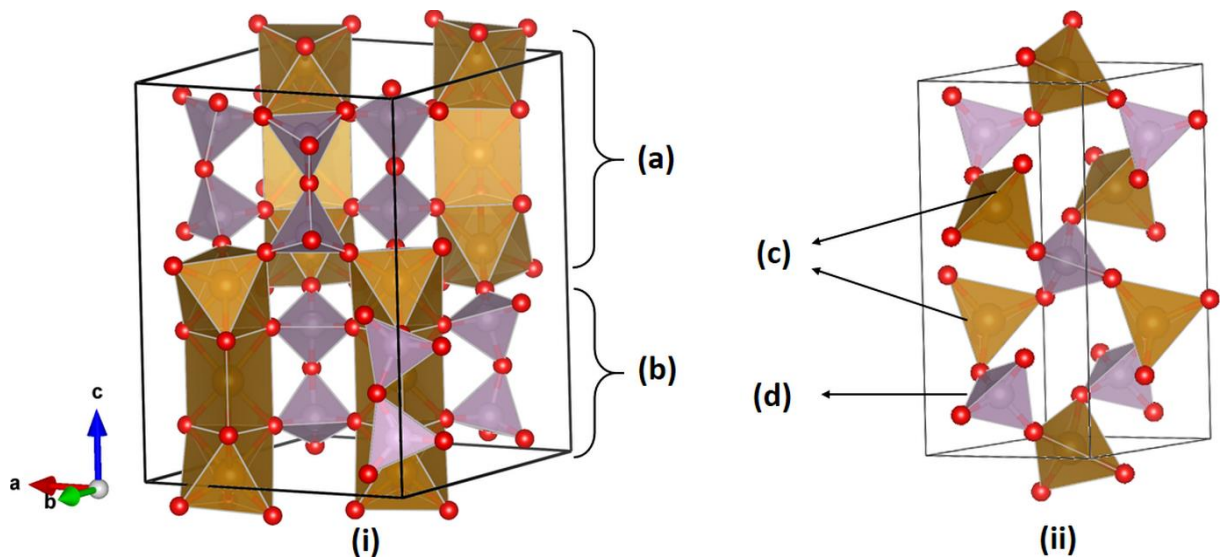

*Figure 8: (i) The $Fe_3(P_2O_7)_2$ crystal showing in (a) $Fe^{2+}$ pyramidal geometry sandwiched between two $Fe^{3+}$ octahedral (b) two $(PO_4)^{3-}$ tetrahedra forming a dimer unit $(P_2O_7)^{4-}$. (ii) $FePO_4$ crystal structure in quartz type tetragonal geometry showing in (c) $Fe^{3+}$ in tetrahedral geometry and in (d) single $(PO_4)^{2-}$ tetrahedron. IPG has a mix of these possibilities in varied proportion depending on preparation conditions.*

**(2)** The way motifs are connected to each other is also complex in IPG since various possible connections are proposed and studied in the literature [50,56,57]. We have developed models incorporating the following connectivity formalisms:

(i) The phosphate tetrahedral units share their oxygen atoms with four units of iron either octahedral or tetrahedral. Similarly, the Iron polyhedral units share their oxygen atoms with phosphate tetrahedral units.

(ii) IPG also have $(P_2O_7)^{4-}$ dimer units (Figure 8(i)(b)). The superior chemical durability of IPG is attributed to the presence of these P-O-P bonds.[57] The percentage of such connections is about 30-40 %. This is also taken care in our modelling.

(iii) $Fe^{2+}$ coordination is not conclusive from any study but there are studies which claim that $Fe_3(P_2O_7)_2$ crystal has most resemblance with IPG structural connections.[57] Therefore, we have modelled $Fe^{2+}$ to be in face sharing distorted geometry with $Fe^{3+}$ octahedron (Figure 8(i)(a)).

(iv) We effectively have $(Fe_3O_{12})^{16-}$ clusters interconnected via $(P_2O_7)^{4-}$ dimer and $(PO_4)^{3-}$ monomer units. The percentages of these units vary according to composition and preparation conditions. In the composition we adopted for modelling (which is reported to be most durable) the following distribution of the units is observed using MD simulations [53,52]:

Fe/P ratio is 0.67 from the formula. P exists mostly in $(PO_4)^{3-}$ tetrahedral form (~100%) due to its strong covalent character. Fe exists in two oxidation states +2 and +3. The ratio of $Fe^{2+}/Fe$ is found to vary from 0 to 20 %. $Fe^{2+}$ exists in distorted octahedral or trigonal prism geometry having face sharing with $Fe^{3+}$ octahedral. $Fe^{3+}$ is found to exist in two geometries i.e. octahedral (75%) and tetrahedral (25%). Hence average coordination number for Fe is between 4 and 6.

The developed structure is shown in Figure 9. The simulation cell is a cube of edge 20Å containing 411 O atoms, 114 P atoms, 78 $Fe^{3+}$ atoms and 9 $Fe^{2+}$ atoms. The structure has 612 atoms. The density of the developed structure is 3.12 gm/cc (experimentally IPG has 2.9 gm/cc). The structure is then compared with the MD developed structure by P. Goj et al. [54] on the basis of structural properties. We also present the structural properties of our structure after force-relaxation using DFT. The need and implications of this is discussed separately later on.

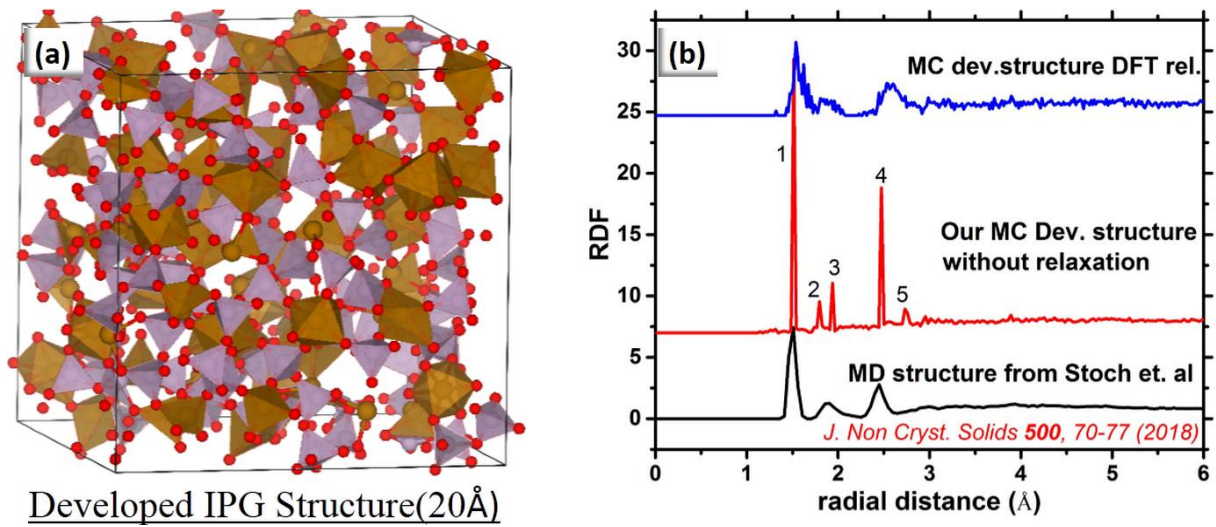

*Figure 9: (a) the developed structure of IPG. For visualisation VESTA software is used. (b) The RDF of the developed structure using MC (in red) is compared with structure developed using MD by P. Goj et al. (in black). Also compared is the total PDF of the structure after DFT relaxation (in blue). Curves are stacked vertically for better clarity.*

In Figure 9 the comparison of total PDF of the structures is also shown. We can see that the PDF produced by our structure matches well with that of the MD developed structure. These peaks are identified with the following bond distances present in the structures. Peak 1 corresponds to P-O bond distance of 1.51 Å present in $PO_4$ tetrahedral motif. Peak 2 and 3 corresponds to Fe-O bond distances of 1.80 Å and 1.94 Å present in tetrahedral and octahedral motifs respectively. Peak 4 and 5 corresponds to O-O bond distance of 2.48 Å and 2.74 Å present in $PO_4$ tetrahedral and $FeO_6$ octahedral motifs respectively. The peak corresponding to O-O bond distance of 2.95 Å present in $FeO_4$ tetrahedral unit is considerably subdued in intensity due to lesser number of such units present in the structure.

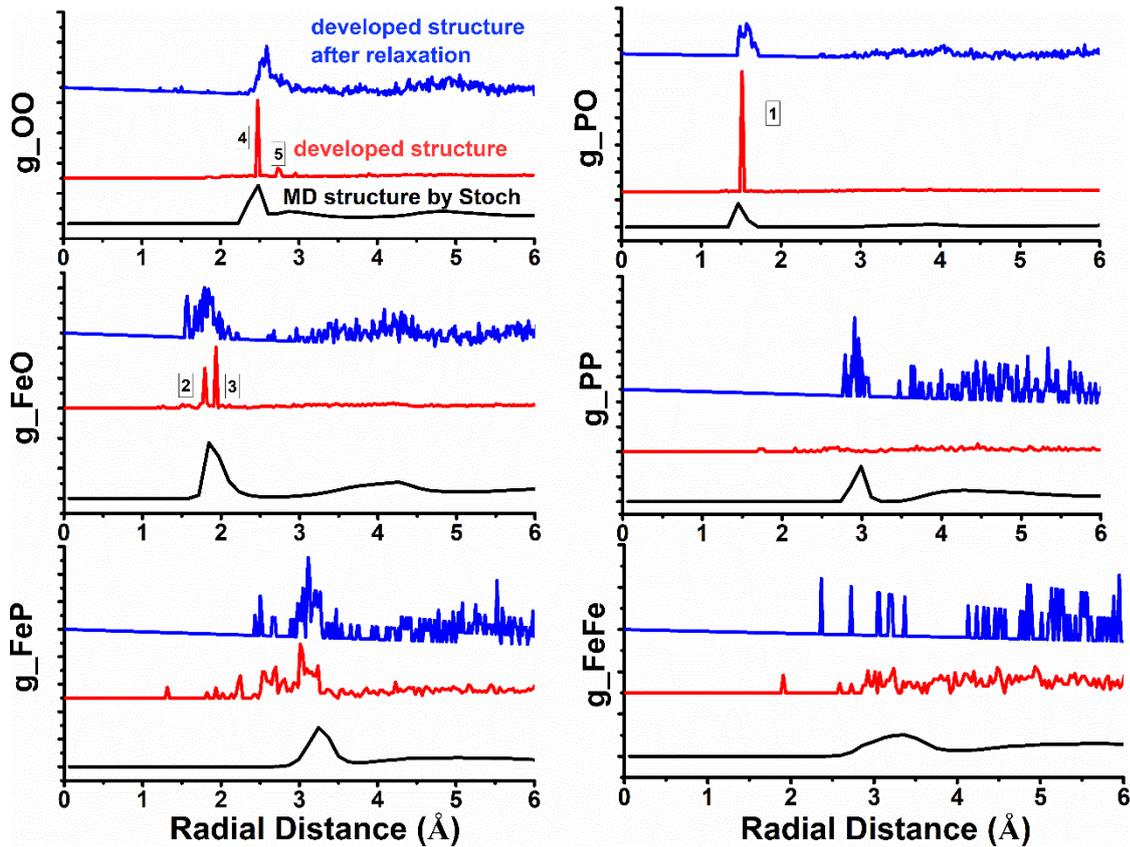

*Figure 10: Graphs showing the comparison of partial pair distribution function of the developed structure before and after relaxation with the MD developed structure by P. Goj et al. Curves are stacked vertically for better clarity*

Figure 10 shows a similar comparison among the structures with respect to partial pair correlation functions. Six possible pair correlation functions in IPG are shown. They give a clearer picture to the peaks discussed in total PDFs. Apart from those peaks, we observe one more peak at 3.27 Å in the Fe-P partial pair correlation function which arises from the Fe octahedral unit and P tetrahedral connection. Apart from Fe-Fe and P-P partial pair correlation function, the agreement is good for the rest of the curves.

The peaks in our case are more sharply defined due to the use of rigid motifs and absence of force-relaxation in our code. The peaks become broader and glass-like after DFT relaxation (observed in all blue PDFs), which is expected. The point to be emphasized here is that though a force relaxation is a must after modelling any glass structure, still the time taken for relaxation of forces can be orders of magnitude smaller if the starting structure is closer to the metastable valley in potential energy landscape (PEL) in which the glass actually resides. Also modelling the key structural aspects through coding helps the relaxation to find the correct metastable valley from the sea of possible valleys in PEL.

**Coordination number analysis**

There are three coordination geometries we are concerned here. First is the average coordination number of oxygen in the structure. For a perfect structure we expect this to be 2, since oxygen acts as bridging atom between Fe and P or between P and P in the network. For our structure this is found to be 2.32 having average P contribution of 1.07 and average Fe contribution of 1.25. The Fe contribution is more due to the presence of edge-sharing distorted polyhedral units with Fe2+ sitting in them. This was designed in the code on the basis of experimental proof regarding the presence of the same. This is found missing in the MD structure which has average coordination number for O as 2.11 with 0.98 contribution from Fe and 1.13 from P. Second is the coordination geometry of P in the structure. Since $(PO_4)^{3-}$ has strong covalent character, ideally the average coordination around P should be 4 (from oxygen). In our structure we have this number to be 3.87 which also compares closely with the value of 3.89 obtained for the MD structure. Finally, the third coordination is that of Fe. From the literature we have information that Fe is present in both tetrahedral and octahedral environments, with $Fe^{2+}$ favouring a distorted octahedral geometry. Thus, the coordination is expected to be less than 6 for Fe in IPG structure due to its ionic character. The value of average coordination number for Fe in our structure is found to be 5.89. The value for MD developed structure is found to be 5.04.

Figure 11 shows the comparison of key bond angles in IPG structure. The O-P-O bond angle distribution gives a peak at 109.5° expected for the $PO_4$ tetrahedral motif. The P-O-P bond angle distribution peaks at ~ 123° in our structure, whereas the MD structure give a peak at ~147°. Upon relaxation our angle distribution peak indeed shifts to right at ~137°. This anomaly can be corrected by placing a constraint on this bond angle in our code, similar to the silica glass case and is addressed in the new models we are generating. The O-Fe-O bond angle distribution in our structure shows three distinct peaks at 90°, 109.5° and 180°. The 90° and 180° peaks are due to the presence of octahedral $FeO_6$ motif with four O-Fe-O angles equal to 90° and one O-Fe-O angle as 180°. This is the reason why the 90° peak has more intensity. Finally, the Fe-O-P bond angle distribution for our structure shows two peaks at ~70° and 140°. The 70° peak is due to the presence of small percentage of Fe in 2+ oxidation state which sit in sterically hindered position between two $Fe^{3+}$ atoms. The 140° peak comes from the usual connection between Fe octahedral or tetrahedral motif with P tetrahedral motif. Upon relaxation the 70° peak is found to relax to higher values and shifts towards right.

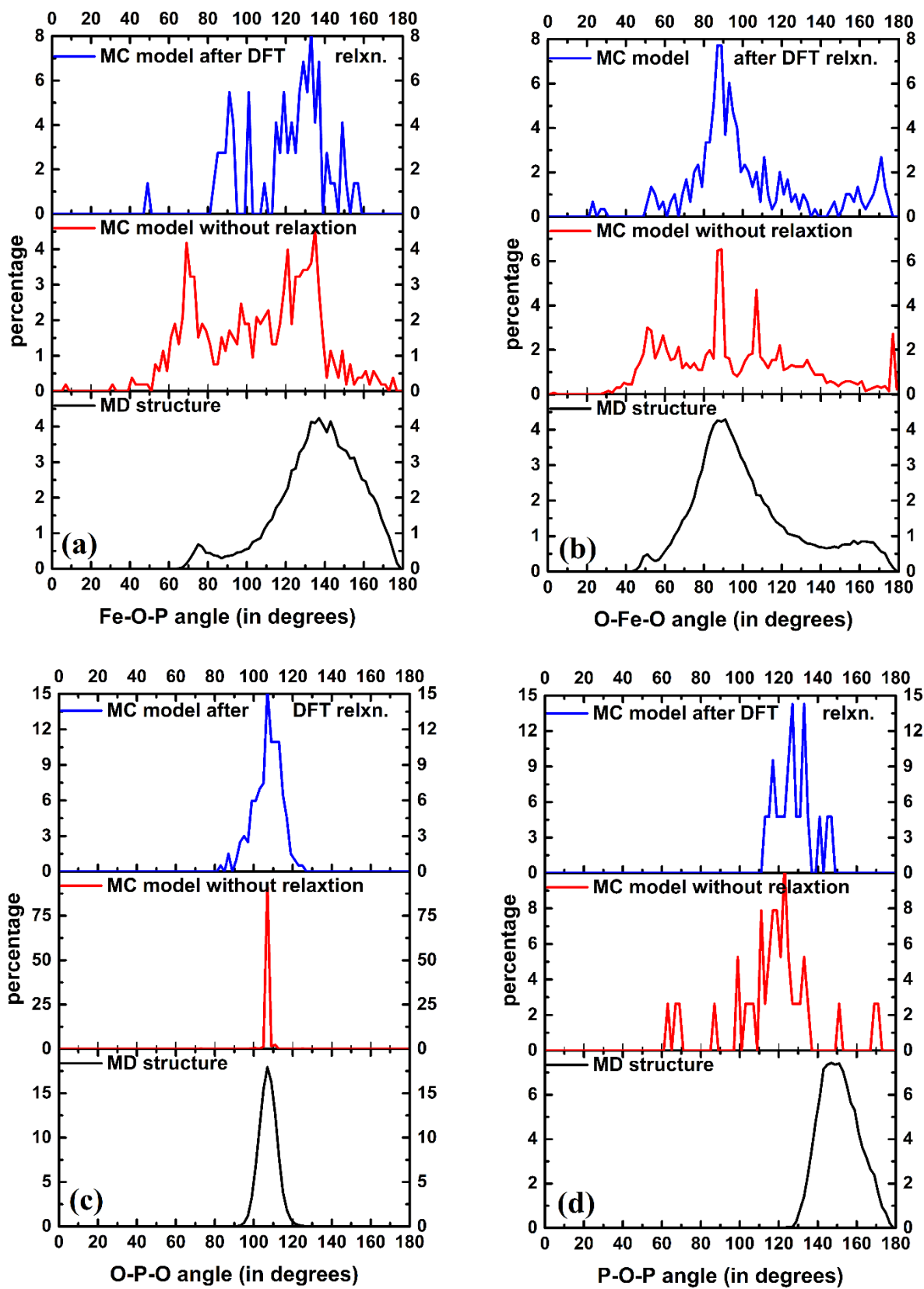

Figure 11: Graphs showing angle distribution in the structure of IPG developed using our code (red) as compared with MD generated structure from P. Goj et al. (black). Also shown is the distribution after our developed structure is relaxed using DFT (blue)

Among the long-range structural properties we found the void fraction and density of the structures. For our structure, the value of void fraction is found to be 0.25 and for the structure by Goj et al. developed using MD its value is 0.27. The density of our structure is found to be 3.12 gm/cc whereas for the MD structure it is found to be 2.92 gm/cc. The experimental density for IPG of this composition is 2.9 gm/cc. The deviation of density of our structures from the experimental value is mainly due to the absence of force relaxation in our structures.

The agreement between the structural properties as developed from our code versus structures developed using MD shows that the code can be applied for modelling complex glass structures as well. Moreover, the topology information from experiments can be designed into the code to produce the required coordination among units. This sets the present method apart from other available methods mentioned earlier.

There are some **drawbacks** of using fixed rigid motifs in modelling glasses as opposed to the real case scenario where the glass motifs are found to have measurable distortion throughout the glass structure. This can be observed from the graphs as well since the PDF peaks in our structures are sharper than RMC or MD developed structure. This shows that the structures developed from our code must be force-relaxed using either MD or DFT prior to using them for further studies. It also implies that our structures can be used as starting structures for glass preparation using DFT, MD or RMC. For validating this point, we developed some structures of IPG with our model and relaxed them using DFT. This is presented next.

For studying the effect of relaxation on energy of these structure, five structures of IPG having composition $40Fe_2O_3:60P_2O_5$ with 0% $Fe^{2+}$ were developed. For force relaxation a strategy appropriate for glasses is adopted where the individual atoms might still be allowed to have some small forces due to the quenched-in disorder but the overall structure should have zero net force. DFT-VASP package is used for carrying out the structural relaxation. Since the structures have around 200 atoms without any symmetry, so only gamma point calculations are carried out with ENCUT=450eV. Thereafter ionic position relaxation and cell volume relaxation were carried out successively many times till net force on all the atoms falls below 0.1 eV/Å.

The result of this exercise is summarised in Table 5. After relaxation the coordination numbers approach to their ideal values (written in brackets). However, ideal values may never be attained in a realistic structure due to the presence of coordination defects in glasses. The effect of relaxation on density and energy of the structure is also recorded. The last column shows the energy per atom gained upon relaxation. The negative values in all the developed structures show the correctness of the relaxation approach adopted for glasses. Also, the data in the table shows the progress of the structure towards the metastable valley in PEL. Hence, the MC modelling and subsequent relaxation of the obtained structure from DFT or MD (depending upon number of atoms involved) is an efficient way to produce glassy structures without bringing melt-quench process into picture. Here only the structural aspect of generated models is discussed, but we are also currently studying the physical and mechanical properties of generated structures using both DFT and MD.

*Table 5: Table showing effect of DFT relaxation of developed structure on Coordination number (CN) of constituents, Density and total energy of the structure; U: unrelaxed, R: relaxed with DFT*

| Developed Structure of IPG | | O CN (2) | P CN (4) | Fe CN (6) | Density (g/cc) (2.9) | Energy change upon DFT relaxation (eV/atom) |
|---|---|---|---|---|---|---|
| Structure 1 | U | 1.74 | 3.41 | 5.52 | 2.94 | -0.16 |
| | R | 1.92 | 3.97 | 5.76 | 3.02 | |
| Structure 2 | U | 1.72 | 3.58 | 5.95 | 2.77 | -0.10 |
| | R | 1.78 | 4.0 | 5.68 | 2.77 | |
| Structure 3 | U | 1.97 | 3.53 | 5.84 | 3.19 | -0.15 |
| | R | 2.07 | 3.86 | 5.92 | 3.29 | |
| Structure 4 | U | 1.83 | 3.42 | 5.96 | 3.16 | -0.11 |
| | R | 1.93 | 3.89 | 5.87 | 3.15 | |
| Structure 5 | U | 1.74 | 3.81 | 5.58 | 2.72 | -0.10 |
| | R | 1.79 | 3.94 | 5.74 | 2.85 | |

# 4. Summary and Conclusion

In summary, a recipe is presented for developing glassy structures which is extensible to complex glasses like IPG and can generate large scale structures utilizing only short-range properties and connectivity information that is apriori available from experiments. We have developed atomistic models of silica and iron phosphate glass using the in-house made MC code. Through a comparison of structural properties from short range to long range with experimental and existing simulated structures, we have shown that using the method presented here it is possible to develop good atomistic models of glasses which incorporate the pre-existing qualitative and quantitative information into the models. The absence of any force calculation in our code makes the development of structures faster than other methods but at the same time leaves for force-relaxation to be done at later stages prior to finding the physical properties using these models. Also, such a method where we can decorate the structure according to pre-existing experimental knowledge can give accurate atomistic models which are not obtainable from either MD (due to lack of correct inter-atomic potential) or DFT (due to computational limits on size of random structure that can be produced). Although the utility of MD and DFT in modelling and studying amorphous systems is immense yet the present shortcomings can be addressed to some extent by starting with nearly correct structures. In addition, the presented method can be used to build a pool of structures for some complex glasses that can further be used to design interatomic potentials for use in MD.

Models of Amorphous Si and Ge. *Phys. Rev. Lett.* **1985**, *54* (13), 1392–1395. https://doi.org/10.1103/PhysRevLett.54.1392.

**Table of Content Graphic**

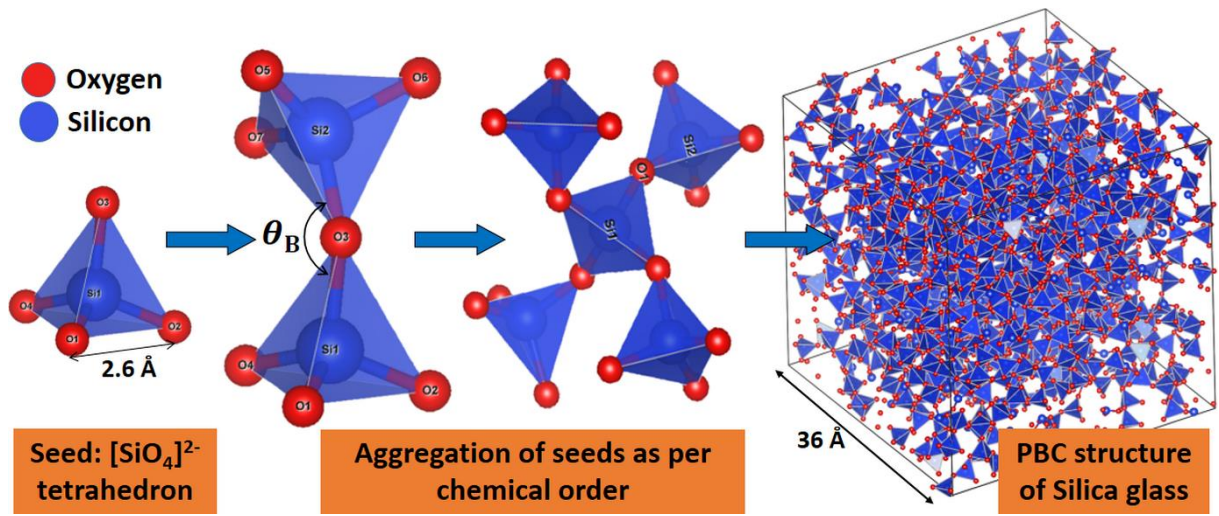